\documentclass[showpacs,prl,twocolumn,aps,floatfix]{revtex4}

 \usepackage{bm}
 \usepackage{amsmath}
 \usepackage{amssymb}
 \usepackage{latexsym}
 \usepackage{amsfonts}
 \usepackage{epsfig}
 \usepackage{psfrag}

 \newcommand{\ket}[1]{\left|#1\right>}
 \newcommand{\bra}[1]{\left<#1\right|}

 \newcommand{\f}[1]{\mbox{\boldmath$#1$}}

 \newcommand{\bea}{\begin{eqnarray}}
 \newcommand{\ea}{\end{eqnarray}}
 \newcommand{\eea}{\end{eqnarray}}
 \newcommand{\ord}{{\cal O}}

\begin{document}

\title{Adiabatic quantum algorithms as quantum phase transitions: 
$1^{\rm st}$ versus $2^{\rm nd}$ order}

\author{Ralf Sch\"utzhold$^*$ and Gernot Schaller}

\affiliation{Institut f\"ur Theoretische Physik,
Technische Universit\"at Dresden, D-01062 Dresden, Germany}

\begin{abstract}
In the continuum limit (large number of qubits), adiabatic quantum
algorithms display a remarkable similarity to sweeps through quantum
phase transitions.
We find that transitions of second or higher order are advantageous in 
comparison to those of first order.
With this insight, we propose a novel adiabatic quantum algorithm for
the solution of 3-satisfiability (3-SAT) problems (exact cover), which
is significantly faster than previous proposals according to numerical
simulations (up to 20 qubits).
These findings suggest that adiabatic quantum algorithms can solve
NP-complete problems such as 3-SAT much faster than the Grover search
routine (yielding a quadratic enhancement), possibly even with an 
exponential speed-up. 
\end{abstract}


\maketitle

The realization that quantum algorithms (e.g., \cite{shor,grover}) can
solve certain problems much faster than (known) classical methods is
one of the main motivations for constructing scalable quantum
computers. 
Unfortunately, these efforts are strongly hampered by the decoherence
induced by the inevitable coupling to the environment, which tends to
destroy the fragile quantum features needed for these quantum
algorithms. 
One idea to overcome this obstacle is adiabatic quantum computation
\cite{farhi} where the solution to the problem to be solved is encoded 
in the ground state of a suitably designed Hamiltonian $H_{\rm out}$.
In order to reach this (unknown) ground state, adiabatic quantum 
algorithms exploit the adiabatic theorem: 
A system described by a time-dependent Hamiltonian $H(t)$ with 
$H(t)\ket{\Psi_n(t)}=E_n(t)\ket{\Psi_n(t)}$
initially prepared in its ground state $\ket{\Psi_0}$ will 
approximately stay in its (instantaneous) ground state -- 
provided the evolution of $H(t)$ is slow enough
$\bra{\Psi_0}\dot H\ket{\Psi_n}\ll(E_n-E_0)^2$.
Starting with an initial Hamiltonian ${H}_{\rm in}$ whose ground 
state is known and easy to prepare, a sufficiently slow evolution into 
$H_{\rm out}$, for example  
\bea
\label{adiabatic-H}
{H}(t)=[1-g(t)]{H}_{\rm in}+g(t){H}_{\rm out}
\,,
\ea
where the parameter $g(t)$ runs from 0 to 1, generates the desired 
final ground state. 
With a sufficiently cold and weakly coupled environment, the occupation 
of the instantaneous ground state should be more robust against the 
impact of decoherence, see, e.g., \cite{robust}.
The adiabatic condition $\bra{\Psi_0}\dot H\ket{\Psi_n}\ll(E_n-E_0)^2$ 
relates spectral properties of ${H}(t)$ with the runtime $T$ 
necessary to obtain a desired probability of the instantaneous ground 
state -- which can be interpreted as the algorithmic complexity of the
quantum algorithm.
However, especially for NP-complete problems such as 3-SAT, the maximum 
speed-up achievable by adiabatic quantum algorithms is still not 
completely clear, see, e.g., \cite{znidaric,farhi-fail,aharonov}.  
In this Letter, we exploit the analogy to quantum phase transitions 
\cite{sachdev} in order to gain new insight into these questions.  

\begin{figure}
\includegraphics[height=4cm]{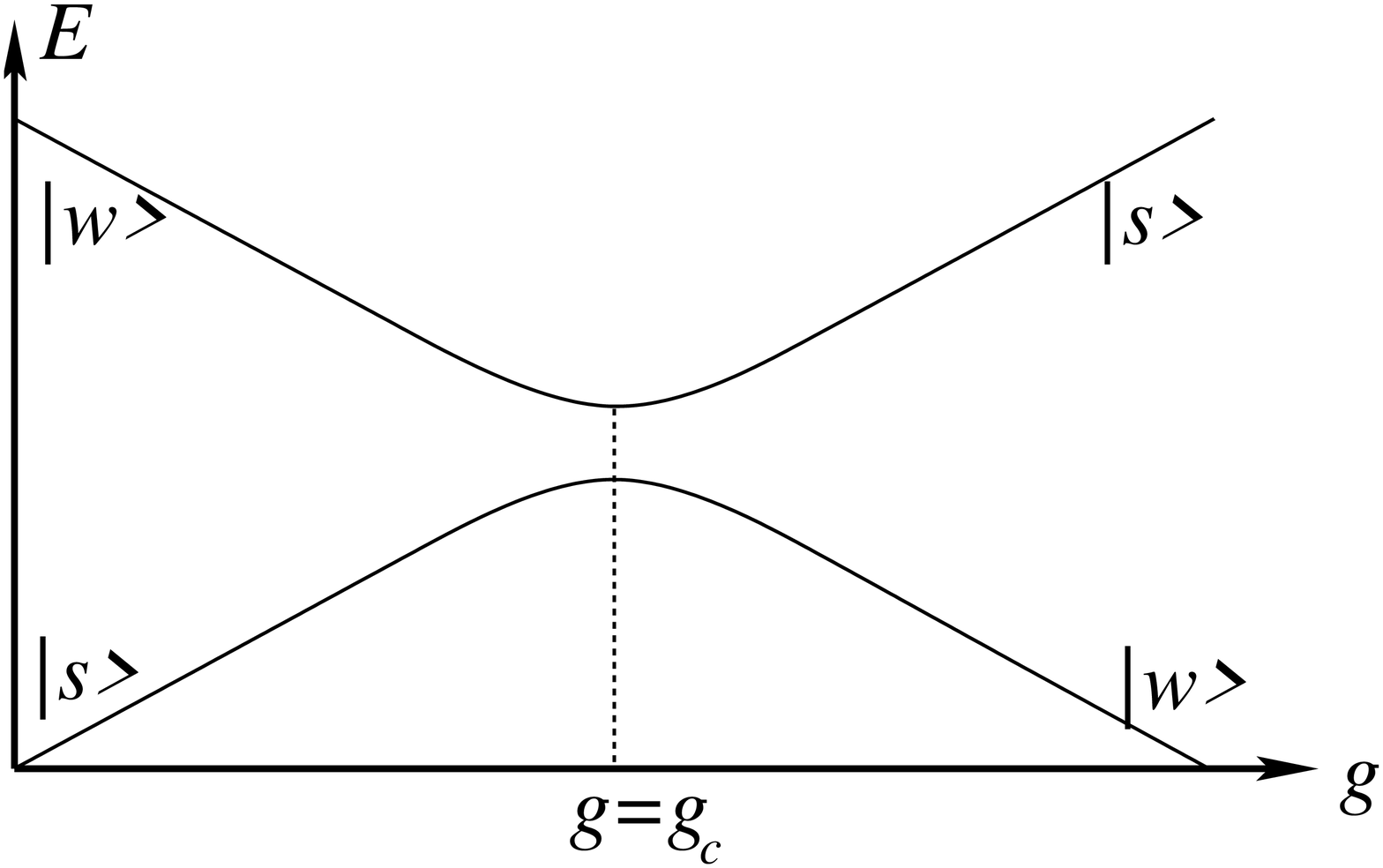}
\caption{\label{Fgrover}Sketch of the level structure 
(ground state and first excited 
  state) of the Grover Hamiltonian in Eq.~(\ref{grover-H}).}
\end{figure}

Let us start with one of the simplest examples: the Grover algorithm, 
which accomplishes the task to find a marked item in an unsorted database 
with $N=2^n$ items with a quadratic speed-up \cite{grover}.
An adiabatic version of Grover's algorithm is defined by the Hamiltonian
\bea
\label{grover-H}
{H}(g)
=
(1-g)\left[\f{1}-\ket{s}\bra{s}\right]+g\left[\f{1}-\ket{w}\bra{w}\right]
\,,
\ea
with $\ket{s}=\sum_{x=0}^{N-1}\ket{x}/\sqrt{N}$ denoting the 
superposition of all numbers from 0 to $N-1$ and $\ket{w}$ 
the marked state, respectively.
Since in this case the commutator between initial and final
Hamiltonians $[H_{\rm in},H_{\rm out}]$ is small, one can nearly 
diagonalize them simultaneously and the $g$-dependent spectrum will
consist of nearly straight lines -- except near $g_{\rm c}=1/2$, where 
we have an avoided level crossing, see Fig.~\ref{Fgrover}. 
In the (continuum) limit of $n\uparrow\infty$, this corresponds to a 
first-order quantum phase transition from 
\mbox{$\ket{s}=\ket{\to\dots\to}$} to
\mbox{$\ket{w}=\ket{\uparrow\downarrow\dots\uparrow\uparrow\downarrow}$} 
at the critical value $g_{\rm c}=1/2$. 
Such a first-order transition is characterized by an abrupt change of the 
ground state ($\ket{s}$ for $g<g_{\rm c}$ and $\ket{w}$ for $g>g_{\rm c}$) 
resulting in a discontinuity of a corresponding order parameter such as 
$\bra{\psi(g)}dH/dg\ket{\psi(g)}=dE/dg$.

\begin{figure}
\includegraphics[height=2.0cm]{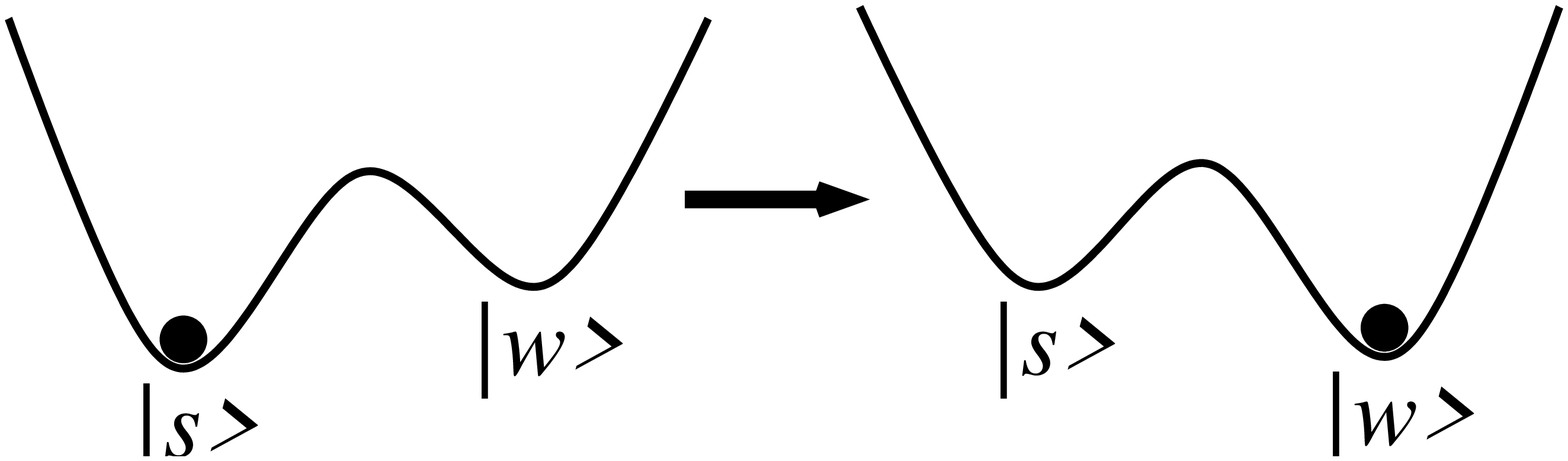}
\caption{\label{Ffirstorder}Sketch of the energy landscape for a 
first-order transition
  with the black dot indicating the ground state.}
\end{figure}

Typically, quantum phase transitions of first order are associated with a 
$g$-dependent energy landscape sketched in Fig.~\ref{Ffirstorder}, where 
the two competing ground states are separated by an energy barrier.
During the phase transition, the system has to tunnel through that barrier 
between $\ket{s}$ and $\ket{w}$ in order to stay in the ground state.
Since naturally the strength of this barrier increases with the system size 
$n$, one would expect the tunneling time to scale exponentially --  
and indeed, the optimal runtime behaves as 
$T=\ord(\sqrt{N})=\ord(2^{n/2})$ for the Grover algorithm described above
\cite{roland}.

\begin{figure}
\includegraphics[height=4.0cm]{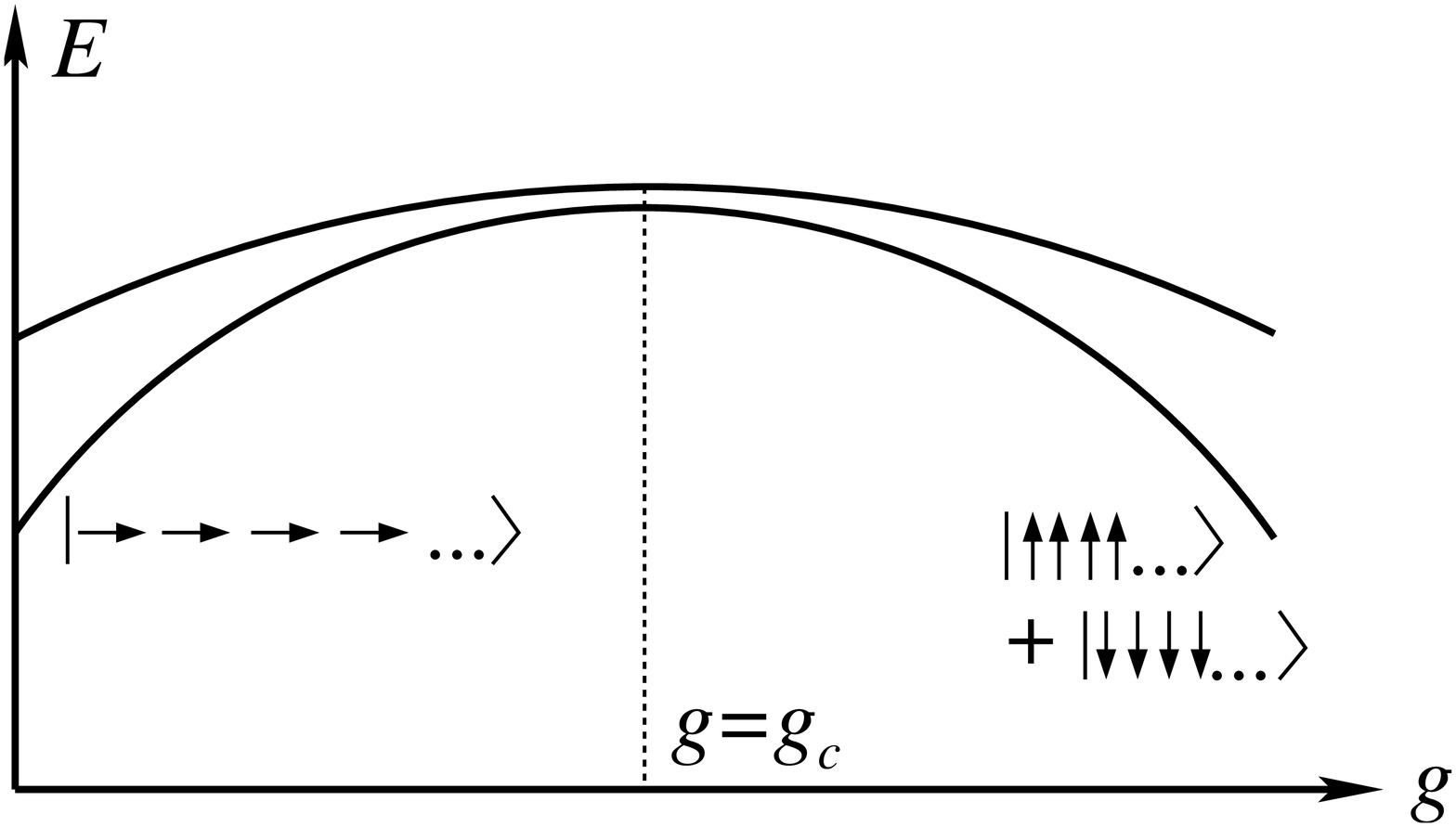}
\includegraphics[height=2.0cm]{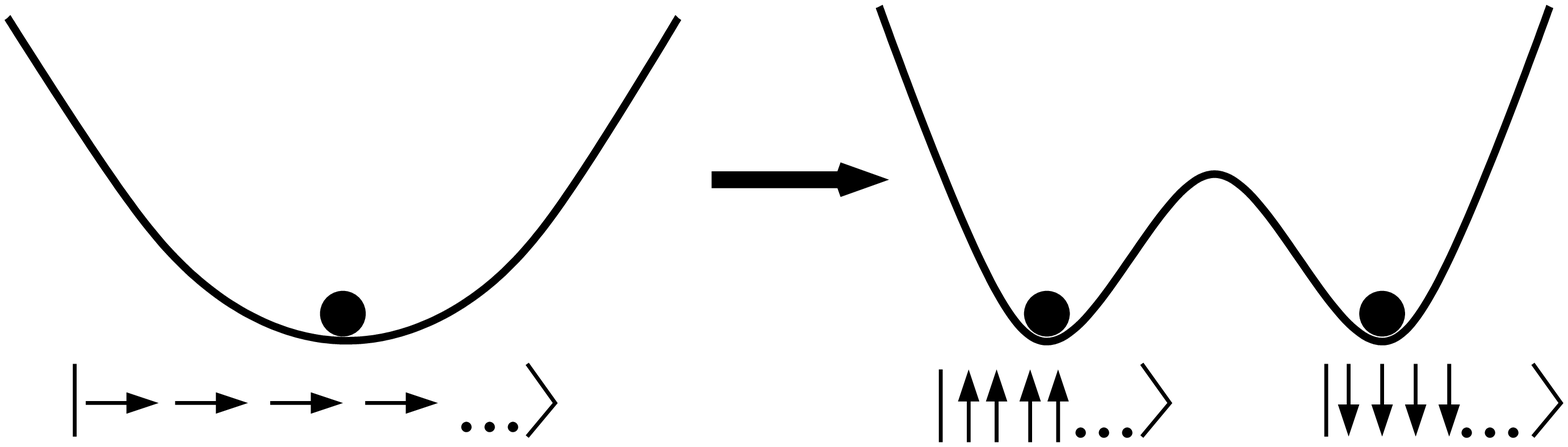}
\caption{Sketch of the level structure (top), i.e., ground state and first 
excited state, and the energy landscape (bottom) of the Ising model as a 
symmetry-breaking second-order phase transition with the black dots 
indicating the unique (left) or degenerate (right) ground state(s).}
\label{Fising}
\end{figure}

The abrupt change of the ground state and the energy barrier suggest that 
first-order transitions are not the best choice for the realization of 
adiabatic quantum algorithms.
In order to study this point in more detail, let us consider a simple 
example for a second-order transition -- the quantum Ising model in the 
presence of a transverse field as defined by the Hamiltonian 
\bea
\label{Eising-H}
{H}(g)
=
-(1-g)\sum\limits_{\alpha=1}^{n}{\sigma}^x_\alpha
-g\sum\limits_{\alpha=1}^{n}{\sigma}^z_\alpha{\sigma}^z_{\alpha+1}
\,,
\ea
where ${\sigma}^{x,y,z}_\alpha$ denote the Pauli matrices acting on 
the $\alpha$th qubit.
This Hamiltonian is invariant under a global 180-degree rotation around the 
$x$-axis which transforms all qubits according to 
\mbox{${\sigma}^z_\alpha\to-{\sigma}^z_\alpha$}. 
The initial ($g=0$) ground state is unique 
\mbox{$\ket{s}=\ket{\to\dots\to}$}, 
whereas the final ground state ($g=1$) becomes two-fold degenerate 
\mbox{$\ket{w_1}=\ket{\uparrow\dots\uparrow}$},
\mbox{$\ket{w_2}=\ket{\downarrow\dots\downarrow}$}, 
and thereby breaks this symmetry.
Typically, such a symmetry-breaking (or restoring) change of the ground 
state corresponds to a second-order phase transition. 
For the Ising model, this expectation can be confirmed analytically by 
an exact diagonalization of the Hamiltonian via Jordan-Wigner and Bogoliubov
transformations (cf.~\cite{sachdev,dziarmaga2005a}).
For such a second-order phase transition, the ground state changes 
continuously (i.e., there is no jump in an order parameter) and the 
energy barrier observed in first-order transitions is absent, 
see Fig.~\ref{Fising}. 
Consequently, one would expect that the system should find its way 
from the initial to the final ground state much easier in this situation -- 
and indeed, the optimal runtime (needed to stay in the ground state) scales 
polynomially in this case since the minimum gap behaves as $\ord(1/n)$, 
cf.~\cite{sachdev,dziarmaga2005a}. 

Note, however, that symmetry-breaking is no guarantee for a second-order
transition. 
As a counter-example (and in order to bring out the difference between 
first and second order transitions a bit more), let us consider another 
simple model Hamiltonian 
\bea
\label{ising+grover}
{H}(g)
=
(1-g)(\f{1}-\ket{s}\bra{s})
-
g\sum\limits_{\alpha=1}^{n}{\sigma}^z_\alpha{\sigma}^z_{\alpha+1}
\,,
\ea
which combines the initial Hamiltonian of the Grover problem with the 
final Hamiltonian of the Ising model.
Even though it has the same symmetry and the same initial and final 
ground states as the Ising model, the above Hamiltonian leads to an 
avoided level crossing corresponding to a first-order transition 
(as can be seen from the local geometry of the spectrum, data not shown 
\cite{preparation}).
As one would expect from the existence of the energy barrier in a 
first-order transition, cf,~Fig.~\ref{Ffirstorder}, 
the scaling of the runtime is exponential in this situation 
\cite{farhi-fail}. 
The main difference between the Ising model and the Hamiltonian above is 
that the latter choice involves $n$-qubit interactions 
$\ket{s}\bra{s}=\prod_{\alpha=1}^n (\f{1}+\sigma^x_\alpha)/2$ 
and therefore the bit structures of the initial and the final 
Hamiltonians are very different. 
These findings may shed additional light onto the recent discussions 
in Refs.~\cite{farhi-fail,znidaric+horvat}.

After having motivated the advantages of second-order phase transitions for
adiabatic quantum computation, let us try to apply these findings to the 
solution of non-trivial computational problems.
In order to compare our results to the literature, we are going to discuss 
a (well studied) special case of a 3-satisfiability problem: exact cover-3.
This problem can be described as follows: a string of $n$ bits 
$z_\alpha\in\{0,1\}$ must satisfy $m$ constraints called clauses.
Each clause involves three bits $\alpha,\beta,\gamma\in\{1, \dots, n\}$ and 
\bea
\label{Eexactcover}
z_\alpha+z_\beta+z_\gamma=1
\ea
is the constraint to be satisfied for every triple 
$(\alpha,\beta,\gamma)$.
Evidently, the solution to this problem, i.e., the bit string satisfying 
all $m$ constraints, is easy to verify but possibly hard to find -- 
i.e., it belongs to the class NP.
It can even be shown that exact cover-3 is NP-complete, i.e., all other 
NP-problems (such as factoring or the traveling salesman problem) 
can be mapped onto exact cover-3 with polynomial overhead. 
Of course, in order to study the speed-up of quantum algorithms in 
comparison with classical methods, we are mostly interested in hard 
instances of this class of problems.
First of all, it is believed \cite{farhi} that problems with a unique 
solution are among the hardest instances of exact cover-3.
A further indication of the complexity can be obtained by treating the 
$m$ constraints of the form given by Eq.~(\ref{Eexactcover}) as a system 
of linear equations.
Clearly, for $m=n$ linearly independent constraints, the solution can be 
found easily.
This suggests that instances with a small number of clauses -- which still 
possess a unique solution -- are particularly hard to solve, see also 
\cite{znidaric}.
We shall consider both cases in the following and compare them. 

In previous approaches (see, e.g., \cite{farhi}), the Hamiltonian 
$H_{\rm out}$ whose ground state encodes the solution to the 
aforementioned problem was constructed by assigning a fixed energy
penalty to each violated clause.
In contrast to this conventional choice 
(which involves three-qubit interactions) we shall use an alternative 
representation requiring two-qubit interactions only.
(In view of an experimental setup, two-qubit interactions are
probably favorable.)
To this end, we sum the terms  
$(\sigma^z_\alpha+\sigma^z_\beta+\sigma^z_\gamma-1)^2/4$ 
over all clauses $(\alpha,\beta,\gamma)$ and obtain \cite{note}
\bea
\label{Efrustrated}
H_{\rm out}
=
\frac14
\sum\limits_{\alpha,\beta=1}^{n}
M_{\alpha\beta}\,\sigma^z_\alpha\sigma^z_\beta
-
\frac12\sum\limits_{\alpha=1}^{n}N_\alpha\,\sigma^z_\alpha
\,,
\ea
plus an irrelevant constant.
Here $N_\alpha$ denotes the number of clauses involving the bit $\alpha$ 
and $M_{\alpha\beta}$ is the number of clauses involving both bits 
$\alpha$ and $\beta$.
The above Hamiltonian corresponds to a frustrated anti-ferromagnet 
in an external field, except that the interaction topology 
$M_{\alpha\beta}$ is defined by the clauses and not by physical 
neighborship.
For satisfiable problems, it has the same ground state (the solution)
as the Hamiltonian used in \cite{farhi}, but some of the excitation
energies differ.

As we have seen in the example in Eq.~(\ref{ising+grover}), the order of 
the phase transition crucially depends on the initial Hamiltonian.
Therefore, the remaining task is to find a suitable initial Hamiltonian 
which respects the bit structure of the final Hamiltonian in 
Eq.~(\ref{Efrustrated}) and whose ground state breaks a global symmetry 
-- which hopefully generates a second-order transition.
The symmetry of the final Hamiltonian we exploit here is its invariance 
under rotations around the ${\Sigma}_z$-axis, where
\bea
\label{Esigma}
{\Sigma}^z=\sum\limits_{\alpha=1}^{n}{\sigma}^z_\alpha
\,.
\eea
An initial Hamiltonian in which this symmetry is spontaneously broken is 
the Heisenberg ferromagnet
\bea
\label{Heisenberg}
{H}_{\rm in} = 
-
\frac14
\sum\limits_{\alpha,\beta=1}^{n}
M_{\alpha\beta}
\,\f{\sigma}_\alpha\cdot\f{\sigma}_\beta
\,.
\eea
Note that both Hamiltonians have the same interaction topology 
$M_{\alpha\beta}$, i.e., bit structure.
In the continuum limit, the ground state manifold of (\ref{Heisenberg})
becomes $SO(3)$-degenerate and contains the separable state
\mbox{$\ket{s}=\ket{\to\dots\to}$}.  
This degeneracy grants us the freedom of choosing the most appropriate 
initial state for the adiabatic algorithm.
The total angular momentum ${\Sigma}_z$ around the $z$-axis is conserved 
during the evolution.
In the final state, ${\Sigma}_z$ counts the difference $\Delta$ between 
the numbers of zeros and ones in the solution (Hamming weight).
Therefore, in order to gain a significant final fidelity, the initial
state should be completely contained in the relevant subspace
\mbox{${\Sigma}_z \ket{\Psi} = \Delta \ket{\Psi}$}.
A suitable initial state  can be generated by the projector 
\bea
\label{projector}
\ket{{\rm in}} = \frac{1}{2n+1} \sum_{k=0}^{2n} 
\exp\left\{2\pi i\,\frac{\Delta - {\Sigma}_z}{2n+1}\,k\right\}
\ket{s}
\,,
\eea
which is just the Fourier decomposition of the Kronecker symbol 
${\delta}(\Delta - {\Sigma}_z)$ and involves single-qubit
rotations only.
Alternatively, one could use an appropriate energy penalty such 
as $({\Sigma}^z-\Delta)^2$, see also \cite{childs2002b}.
Of course, for this initial state preparation, we have to know $\Delta$.
However, this is not a major obstacle: we have found that for the hard 
instances we consider (see results), the number of ones in the solution 
is sharply peaked around $\Delta=n/3$. 
In any case, the overhead of trying every possible value of $\Delta$ 
scales linear (i.e., polynomial) in $n$. 

Numerically, we found that an initial Hamiltonian corresponding to 
the transversal $x,y$-ferromagnet 
\bea
\label{Eferromagnet}
{H}_{\rm in} = 
-
\frac14
\sum\limits_{\alpha,\beta=1}^{n}
M_{\alpha\beta}
\left(
\sigma^x_\alpha\sigma^x_\beta
+
\sigma^y_\alpha\sigma^y_\beta
\right)
\,,
\eea
yields an even better performance than the one in Eq.~(\ref{Heisenberg}). 
In this case, the exact $SO(3)$-degeneracy of the ground state of 
Eq.~(\ref{Heisenberg}) is replaced by an approximate $O(2)$-degeneracy 
(mean-field approximation) generated by ${\Sigma}_z$, i.e., the state 
in Eq.~(\ref{projector}) has a large overlap with the exact ground state 
of Eq.~(\ref{Eferromagnet}) in the relevant ${\Sigma}_z$-subspace. 

In order to test the performance of the linear interpolation between 
the Hamiltonians (\ref{Eferromagnet}) and (\ref{Efrustrated})
proposed here and to compare it with the conventional interpolation 
scheme used in \cite{farhi}, for example, we have simulated the adiabatic 
quantum algorithms numerically \cite{numerics}.  
For different qubit numbers $n$, we have randomly generated instances 
of the exact cover-3 problem with a unique solution in complete analogy
to the procedure described in \cite{farhi}. 
In addition -- as motivated by the comments after Eq.~(\ref{Eexactcover}) 
and Ref.~\cite{znidaric} -- we have also generated hard subsets of uniquely 
satisfying agreements with especially few clauses. 
Technically, this was done by keeping those instances that had 
$m \le {\rm round}(2n/3)$ clauses only.

\begin{figure}
\centerline{\mbox{\epsfxsize=8cm\epsffile{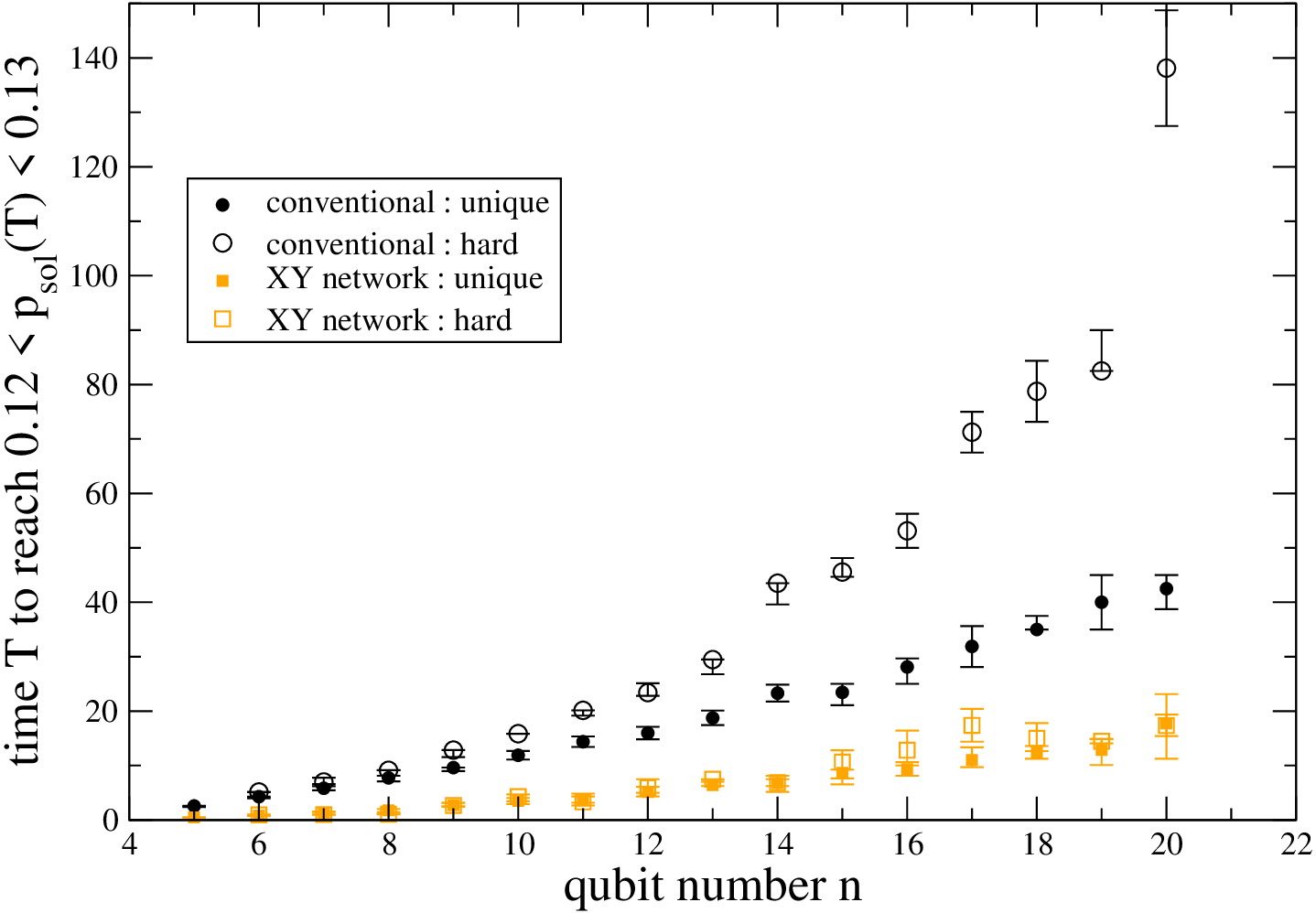}}}
\caption{\label{Fruntimes} Runtime $T$ necessary to reach final 
fidelity of 1/8. Data points show the median out of 100 instances, 
whereas error bars display the 95\% confidence interval on the median.} 
\end{figure}

The results of our numerical simulations are presented in 
Fig.~\ref{Fruntimes}.
For the conventional interpolation scheme \cite{farhi} applied to 
randomly generated instances admitting unique solutions, 
we reproduce the results known from the literature:
Fits to the median runtime yield the same quadratic scaling as in 
\cite{farhi} and for the corresponding minimum fundamental gap 
we obtain similar results as in \cite{latorre} 
(not shown, \cite{preparation}).
However, it becomes visible that for the instances with few clauses 
(which we believe to be very hard), the performance of the 
conventional scheme \cite{farhi} deteriorates significantly.
This is consistent with the observation that the minimum gap is
considerably smaller for those hard instances than for the other
instances with more clauses in the conventional interpolation scheme 
(data not shown \cite{preparation}).

In comparison, the novel adiabatic quantum algorithm ($x,y$-network) 
based on a linear interpolation between (\ref{Eferromagnet}) and 
(\ref{Efrustrated}) proposed here yields a superior performance and 
scaling behavior up to the range of $n=20$ qubits -- 
which becomes even more pronounced for the hard instances.
Note that the data points for those hard instances cluster around 
qubit numbers that can be divided by three, which is probably a 
consequence of our restriction $m \le {\rm round}(2n/3)$.
Unfortunately, the error bars and the small problem sizes (albeit at
the limit of our computational abilities) do not allow to draw
conclusions whether the limiting scaling is exponentially or
polynomially for this NP-complete problem.
However, our results strongly indicate that the scaling is better than 
that of the Grover search routine with $T=\ord(\sqrt{N})$.
Although these results are encouraging, it should be stressed that 
the average behavior (median) can be quite different from the worst 
case scenario.
Indeed we did also encounter instances for which the required runtime
was significantly longer and the associated gap was very small (the
median is insensitive to these).
In many of these extremal cases, our novel algorithm was still superior,  
but sometimes the conventional scheme \cite{farhi} performed better on
these instances. 

In summary, the analogy to quantum phase transitions facilitates a better 
understanding of adiabatic quantum algorithms.
Apart from the vanishing energy gap at the critical point 
(in the continuum limit), another typical signature for the occurrence 
of a phase transition is the divergence of the entanglement, 
cf.~\cite{latorre}.
The energy barrier occurring in first-order transitions provides an 
intuitive explanation for the exponential scaling of the runtime observed 
in these situations. 
The absence of this barrier in transitions of second (or higher) order 
gives raise to the hope that suitably designed adiabatic quantum 
algorithms might yield a much better scaling behavior -- possibly even 
an exponential speed-up.

This work was supported by the Emmy Noether Programme of the 
German Research Foundation (DFG) under grant No.~SCHU~1557/1-1/2.
R.~S.~acknowledges fruitful discussions at the 
Les Houches Summer School on Quantum Magnetism (supported by PITP), 
the Workshop ``Low dimensional Systems in Quantum Optics'' at the 
CIC in Cuernavaca (supported by the Humboldt foundation), 
and valuable conversations with G.~Volovik 
(visits supported by EU-ULTI and ESF-COSLAB). 
The authors are indebted to F.~Krauss for providing computational
resources and to E.~Farhi and J.~Goldstone for helpful comments.
\\
$^*$\,{\small\sf schuetz@theory.phy.tu-dresden.de}

\end{document}